\documentstyle[12pt]{article}
\newcommand{\eul}[1]{{#1}}
\newcommand{\reg}{\mbox{regular}}
\newcommand{\you}{\widehat{{u}}(1)}
\newcommand{\ope}[2]{\frac{#2}{(z-w)^#1}}
\newcommand{\opeone}[1]{\frac{#1}{z-w}}
\newcommand{\aff}[1]{\widehat{{#1}}}

\begin{document}
\begin{titlepage}
\begin{center}
           \hfill    Preprint-KUL-TF-91/33\\

\vskip .5in

{\LARGE {\sc $W_3$ constructions\\ on$_{}$
 \\ affine Lie algebras}}

\vskip .5in

{\large Alex Deckmyn\footnote{Aspirant N.F.W.O. Belgium, e-mail:
fgbda23@blekul11.BITNET} and
Stany Schrans\footnote{Onderzoeker, I.I.K.W. Belgium, e-mail:
fgbda31@blekul11.BITNET}\\ Instituut voor
Theoretische Fysica\\K.U. Leuven, Celestijnenlaan 200-D\\
B-3001 Leuven\\BELGIUM
\vskip .5in
\date{August 1991}}
\end{center}
\begin{abstract}
We use an argument of Romans showing that every Virasoro construction
leads to realizations of $W_3$, to
construct $W_3$ realizations on arbitrary affine Lie
algebras. Solutions are presented for generic values of the level as well as
for
specific values of the level but with arbitrary parameters. We give a detailed
discussion of the $\aff{su}(2)_\ell$-case. Finally, we
discuss possible applications of these realizations to the construction of
$W$-strings.
\end{abstract}
\end{titlepage}
\newpage
\setcounter{page}{1}
\setcounter{footnote}{0}

\noindent {\bf 1.}  It has become clear in the last few years that
$W$-algebras---nonlinear extensions of the Virasoro algebra by
additional primary fields of dimension greater than two---play an important
r\^{o}le in two-dimensional
conformal field theory as well as in integrable systems.  These algebras have
recently been gauged, giving the so-called $W$-gravity theories.

In order to study the representation theory of these algebras, it is convenient
to realize them in terms of simpler structures, such as free fields or affine
currents.  The interest in realizations of  conformal field theories on
affine algebras has increased considerably since it has been noted that the
Sugawara and coset energy-momentum tensors are but specific examples of far
more
general ``Virasoro constructions". The most general realization of a Virasoro
algebra  as a quadratic combination of affine currents (including possible
background charges) leads to a set of coupled algebraic equations
\cite{halpkit,moretal}. Though only quadratic, these ``master  equations" have
so far  resisted any attempt at a general solution. However, the use of
specific ans\"{a}tze has led to numerous new solutions,  lifting  a tip of
the veil on this  space of constructions.  See {\em e.g.}
\cite{irrcons,schratro} for some
of these new solutions and early developments.

For $W$-algebras, and in particular for its simplest example, $W_3$, an
analogous approach has been unsuccessful up to now. In fact,
the straightforward albeit tedious task of writing  down the corresponding
``master equations" for $W_3$---and thus generalizing the coset  solutions of
\cite{bbss}---still remains to be done. So far, the most general results
have been obtained in \cite{romans} for $W_3$, using $N$ free scalar
fields---{\em i.e.} an abelian affine algebra $(\you)^N$.

In this letter, we construct realizations of $W_3$ for arbitrary affine
Lie (super)algebras. The starting point is an argument by Romans in
\cite{romans} which reduces the problem of $W_3$ constructions to the one of
Virasoro constructions. Although this certainly does not give the most  general
$W_3$ construction, it allows us to associate many different such constructions
with every affine algebra.  We  illustrate this method with various affine
algebras, yielding constructions for generic values of the level as well as for
specific values of the level but with free parameters. In particular, we write
down explicitly a realization of $W_3$ in terms of $\aff{su}(2)$ currents for
generic value of the level.
Also, we show how the $c=2$ free field realization of
$W_3$ from \cite{fazam} can be recast as an affine one-parameter construction
on $\aff{su}(2)_4\times \you$. Finally, we discuss possible applications of
these
realizations to the constructions of $W$-strings.

\vspace{.3cm}

\noindent {\bf 2.}
The $W_3$ algebra \cite{zamo} is an extension of the Virasoro algebra,
which is generated
by an energy-momentum tensor $T(z)$ satisfying
\begin{equation}
T(z)T(w)= \ope{4}{c/2}+\ope{2}{2T(w)}+\opeone{\partial T(w)}+\reg,
\label{eq:vir}
\end{equation}
and a dimension 3 primary field $W(z)$ with
\begin{equation}
T(z)W(w)= \ope{2}{3W(w)}+\opeone{\partial W(w)}+\reg .
\end{equation}
The $W(z)W(w)$ operator product expansion is more complicated; it is given by
\begin{eqnarray}
&&W(z)W(w)=\ope{6}{c/3} +\ope{4}{2T(w)}+\ope{3}{\partial T(w)}\nonumber\\
&&\;\;\;\;+
\ope{2}{\frac{32}{5c+22}\Lambda(w)+\frac{3}{10}\partial^2T(w)}
+\opeone{\frac{16}{5c+22}\partial
\Lambda(w)+\frac{1}{15}\partial^3T(w)}
+\reg , \nonumber\\
&& \label{eq:w3ope}
\end{eqnarray}
where we have introduced the composite field
\begin{equation}
\Lambda (z) =(TT)(z)-\frac{3}{10}\partial ^2 T(z).\label{eq:ealambda}
\end{equation}
As usual, the brackets around operators denote normal ordering defined by
point-splitting regularization.

Starting from an arbitrary conformal field theory ${\cal A}$,
whose energy-momen\-tum tensor  $T_0(z)$ generates a Virasoro algebra
(\ref{eq:vir}) with central charge $c_0$, one can construct a realization of
$W_3$ in the following way \cite{romans}. One adds a
$\you$ current $J(z)$, commuting with $T_0(z)$, with operator product expansion
\begin{equation}
J(z)J(w)=\ope{2}{1}+\reg,
\end{equation}
and an energy-momentum tensor
\begin{equation}
T_{\you}(z) = \frac{1}{2} (JJ)(z) +\frac{1}{2}\sqrt{1-c_0}
\partial J(z) \label{eq:tu1},
\end{equation}
generating a Virasoro algebra with central charge $3c_0-2$.
The energy-momentum tensor for ${\cal A} \times \you$ is then given by
$T(z)=T_0(z)+T_{\you} (z)$ and generates a Virasoro algebra with central charge
\begin{equation}
c=2(2c_0-1).\label{eq:cw3}
\end{equation}
It is then a straightforward exercise to check that\footnote{We
have done this using the {\em Mathematica$^{TM}$} package {\sl OPEdefs}
\cite{kris}.}
\begin{eqnarray}
W(z)&=&-\frac{2\sqrt{2}}{3\sqrt{5c+22}} \left[ (J(JJ))(z)
+\frac{3}{2}\sqrt{1-c_0} (J\partial J)(z)\right. \nonumber\\
&&\left.+\frac{1-c_0}{4} \partial^2J(z)
-6(JT_0)(z)-\frac{3}{2}\sqrt{1-c_0} \partial T_0(z)   \right] \label{eq:wfield}
\end{eqnarray}
is a dimension 3 primary field w.r.t. $T(z)$ and generates a $W_3$ algebra on
${\cal A}\times \you$.
In fact, one can explicitly check that another choice of coefficients  in
(\ref{eq:tu1}) and (\ref{eq:wfield}) does not lead to a $W_3$ algebra for
generic $c_0$.

The power of this method resides in the fact that it reduces the search for
generalized $W_3$  constructions to the one for generalized Virasoro
constructions; the latter one being much simpler\footnote{Of course not all
$W_3$ realizations are of this  form, see {\em e.g.} the coset construction of
\cite{bbss}.}. In fact, as mentioned in the introduction, such generalized
Virasoro constructions have recently received a lot of attention in the context
of affine Lie algebras. Every such generalized Virasoro construction on an
affine Lie algebra $\widehat{\eul{g}}$ hence automatically provides a
realization of $W_3$ on $\widehat{\eul{g}}\times \you$, but in some cases  also
on $\aff{g}$, as we will explain shortly.

Notice that this construction does not display a ``conjugation" property
analogous to the one for the Virasoro construction on affine Lie algebras
\cite{halpkit},  where the energy-momentum tensors generating a Virasoro
algebra come in pairs. Given any such energy-momentum tensor $T(z)$,  one can
of course construct a $W_3$ algebra starting from  this tensor or from its
conjugate partner $\tilde{T}(z)=T_{Sugawara}(z)-T(z)$. However, since the
Sugawara tensor has no ``natural" generalization to the $W_3$ algebra
\cite{bbss}, there is no straightforward
generalization of this conjugation property to $W_3$.

Let us illustrate the above construction with some examples.

\vspace{.3cm}

\noindent {\bf 3.}
The standard Virasoro
construction on a simple affine algebra $\widehat{\eul{g}}$ at level $\ell$
is the Sugawara construction with central charge $c_0=\ell \mbox{dim}\,
\eul{g}/(\ell+h_\eul{g})$, where $h_\eul{g}$ denotes the dual coxeter number
of $\eul{g}$.
This naturally leads to a $W_3$
construction on
$\widehat{\eul{g}}_{\ell}\times \you$ with central charge
\begin{equation}
c= \frac{2[\ell (2 \mbox{dim}\,\eul{g}-1)-h_\eul{g}]}{\ell+h_\eul{g}}.
\label{eq:cvalues}
\end{equation}
The generalization to semi-simple algebras is straightforward.
These values are summarized for the simple algebras in table 1.

\begin{center}
\begin{tabular}{|c|c|} \hline
algebra &central charge\\
\hline
$\widehat{\eul{su}}(n)_{\ell}\times \you$\\
$\widehat{\eul{so}}(2n)_{\ell}\times \you $&$
\frac{2[\ell (4n^2-2n-1)-2n+2]}{\ell+2n-2}$\\
$\widehat{\eul{so}}(2n+1)_{\ell}\times \you $&$
\frac{2[\ell (4n^2+2n-1)-2n+1]}{\ell+2n-1}$\\
$\widehat{\eul{sp}}(2n)_{\ell}\times \you $&$
\frac{2[\ell (4n^2+2n-1)-n-1]}{\ell+n+1}$\\
$(\widehat{G}_2)_{\ell}\times \you $&$
\frac{2(27\ell-4)}{\ell+4}$\\
$(\widehat{F}_4)_{\ell}\times \you $&$
\frac{2(103\ell-9)}{\ell+9}$\\
$(\widehat{E}_6)_{\ell}\times \you $&$
\frac{2(155\ell-12)}{\ell+12}$\\
$(\widehat{E}_7)_{\ell}\times \you $&$
\frac{2(265\ell-18)}{\ell+18}$\\
$(\widehat{E}_8)_{\ell}\times \you $&$
\frac{2(495\ell-30)}{\ell+30}$\\
\hline
\end{tabular}\\
\underline{table 1}: $W_3$-construction on Sugawara\\
$\;\;\;\;\;\;\;\;\;\widehat{\eul{g}}_{\ell}\times \you$ for simple $\eul{g}$
\end{center}

The method can of course be applied to more general Virasoro
constructions than the Sugawara one: it holds in particular for any of the
coset energy-momentum tensors and for any of the new irrational Virasoro
constructions on affine Lie algebras---see {\em e.g.}
\cite{irrcons,schratro}---or even
affine Lie superalgebras \cite{godolwat,liesuper}, and also for the different
one-parameter solutions at fixed level with fixed value of the central charge
\cite{irrcons,schratro,liesuper}.

\vspace{.3cm}

\noindent {\bf 4.}
In fact, the method is not restricted to $\aff{g}\times \you$, but can be
applied to any affine $\aff{g}$. Indeed, the commuting $\you$ current may
sometimes be taken in $\aff{g}$ itself. Let us illustrate this with the coset
construction.

Suppose that $\eul{g}$ has a subalgebra $\eul{h}$ and consider the coset
energy-momen\-tum tensor $T_{\eul{g}/\eul{h}}(z)=  T_\eul{g}(z)-T_\eul{h}(z)$
where $T_\eul{g}(z)$ and $T_\eul{h}(z)$  are the Sugawara energy-momentum
tensors for $\aff{g}$ and $\aff{h}$, respectively.  Any current $J(z)$ in
$\aff{h}$
commutes automatically with  $T_{\eul{g}/\eul{h}}(z)$, leading to a
construction on $\aff{g}$. This is rather surprising since
it means that a $W_3$ algebra can be constructed using currents of any
arbitrary affine Lie (super)algebra, provided it has a Sugawara energy-momentum
tensor! Consider as an example $\aff{g}=\aff{su}(2)$ with
arbitrary level $\ell$.   The defining operator product expansion can be
written as
\begin{equation}
J^i (z)J^j (w) = \ope{2}{\ell \delta^{ij}/2}+
\opeone{i \epsilon ^{ijk} J^k (w)}
+\reg, \label{eq:su2ope}
\end{equation}
where $\epsilon ^{ijk}$ is completely antisymmetric and $\epsilon ^{123}=1$.
Take as subalgebra the $\you$ generated by $J^3(z)$, leading to the coset
energy-momentum tensor
\begin{equation}
T_0(z)=T_{\eul{su}(2)/\eul{u}(1)}(z)=\frac{1}{\ell+2}\left((J^1J^1)(z)+
(J^2J^2)(z)\right) -\frac{2}{\ell(\ell+2)}(J^3J^3)(z)
\end{equation}
with central charge $c_0=2(\ell-1)/(\ell+2)$. Setting $J(z)=
\sqrt{\frac{2}{\ell}}J^3(z)$ then leads to a
$W_3$ algebra on $\aff{su}(2)_\ell$ generated by
\begin{eqnarray}
T(z)&=& \frac{1}{\ell+2}\left((J^1J^1)(z)+(J^2J^2)(z)+(J^3J^3)(z)
\right)+\sqrt{\frac{4-\ell}{2\ell(\ell+2)}} \partial J^3(z)  ,\nonumber\\
W(z) &=& \sqrt{\frac{2}{(13\ell-4)(\ell+2)}}\left[ 2\sqrt{\frac{2}{\ell}}\left(
(J^3(J^1J^1))(z)+(J^3(J^2J^2))(z)\right)\right.\nonumber\\
&&-\frac{2(\ell+8)}{3\ell}\sqrt{
\frac{2}{\ell}}(J^3(J^3J^3))(z)+\sqrt{\frac{4-\ell}{\ell+2}}\left( (J^1\partial
J^1)(z) + (J^2\partial J^2)(z)\right)\nonumber\\
&&\left.-\frac{\ell+4}{\ell}\sqrt{\frac{4-\ell}{\ell+2}}(J^3\partial J^3)(z)
-\frac{4-\ell}{12} \sqrt{\frac{2}{\ell}}\partial^2 J^3\right]
\end{eqnarray}
with central charge
\begin{equation}
c=6\frac{\ell-2}{\ell+2}.
\end{equation}
For $\ell=4$ ($c=2$) this realization becomes extremely simple: $T(z)$ is
then the Sugawara tensor, while the noncubic terms in $W(z)$ vanish.

It was pointed out in \cite{schratro} that a number of the new
irrational $\aff{su}(3)$ Virasoro
constructions commute with one of the $\aff{su}(3)$ currents.
There too, an additional $\you$ current is not needed.

\vspace{.3cm}

\noindent {\bf 5.}
When the conformal field theory ${\cal A}$ has central charge $c_0=1$, then
(\ref{eq:tu1}) and
(\ref{eq:wfield}) become particularly simple, since the square
roots vanish in that case. A nice example is provided by the one-parameter
Virasoro construction for $\widehat{\eul{su}}(2)$ at level 4 of \cite{moretal}
which has $c_0=1$.
Adding a commuting $\you$ current $J^0(z)=\sqrt{2} J(z)$
to the algebra (\ref{eq:su2ope})
with $\ell=4$, the $W_3$ realization corresponding to the one-parameter
$\aff{su}(2)_4^\#$ solution  can be written as
\begin{eqnarray}
T(z)&=&\sum_{\mu=0}^3 \lambda^\mu (J^\mu J^\mu)(z),\nonumber\\
\sqrt{2} W(z)&=&-\frac{1}{3}(J^0(J^0J^0))(z) + \sum_{\mu=0}^3
\lambda^\mu (J^0(J^\mu J^\mu))(z) \label{eq:c2par},
\end{eqnarray}
where the $\lambda^\mu$ satisfy
\begin{equation}
\sum_{i=1}^3 \lambda^i= \lambda^0 =\frac{1}{4}
\end{equation}
and
\begin{equation}
\sum_{i=1}^3 (\lambda^i)^2 = \frac{1}{16}.
\end{equation}
They are thus determined up to an arbitrary parameter $\theta$ and can be
written explicitly as $\lambda^0=\frac{1}{4}$ and
\begin{eqnarray}
\lambda^1 &=& - \frac{1}{3}\cos\frac{\theta}{2}\cos\left(
\frac{\theta}{2} + \frac{2\pi}{3}\right),\nonumber\\
\lambda^2 &=& - \frac{1}{3}\cos\left(\frac{\theta}{2} +
\frac{2\pi} {3}\right)\cos\left(\frac{\theta}{2} +
\frac{4\pi}{3}\right),\nonumber\\
\lambda^3 &=& - \frac{1}{3}\cos\left(\frac{\theta}{2} +
\frac{4\pi} {3}\right)\cos\frac{\theta}{2}.
\end{eqnarray}

This algebra has central charge equal to two. It has been shown that the
one-parameter Virasoro construction $\widehat{\eul{su}}(2)_4^\#$ has a
realization in terms of a free scalar field compactified on a circle
\cite{morshiftur}. Plugging this realization in  (\ref{eq:c2par}) yields
precisely the free field realization at $c=2$ of \cite{fazam}. A natural
question to ask at this point is whether this generalizes to the other $W_n$
algebras \cite{faly}, {\em i.e.} does the (Miura) free field realization of
$W_n$,  at $c=n-1$  have a similar  interpretation of a $W_n$ construction
on a nonabelian affine Lie algebra?

\vspace{.3cm}

\noindent {\bf 6.} A classical version of the $W_3$ algebra can be gauged to
give a $W_3$ gravity theory \cite{hullletter}. The corresponding quantum
theory is in general anomalous \cite{anomgrav}.  The universal anomalies vanish
if the matter system has critical charge $c_{crit}=100$ \cite{mieg}, while
matter anomalies are related to the fact that a realization of the classical
$W_3$  algebra is not necessarily a realization of the quantum $W_3$ algebra
(the one considered here), and depend on the explicit form of the realization.
Any (quantum) realization of $W_3$ with $c=100$ is thus a candidate to
construct anomaly-free $W_3$ gravity, and hence $W_3$ string theory.
Taking into account four spacetime dimensions, this leads to a conformal field
theory ${\cal A}$ with central charge $c_0=49/2$.
A little experimenting with the $c$-values of table 1, reveals a considerable
number of possibilities, most of which are, unfortunately, quite complicated.
The simplest example on Sugawara $\aff{g}_l\times \you$ is
$\aff{g}_\ell=\aff{so}(49)_1$, while $\frac{\aff{so}(51)_1}{\you} \times \you$
and $\frac{\aff{su}(1,1|7)_{85}}{\you}\times \you$
give examples\footnote{See \cite{godolwat} for an account of the Sugawara
construction on affine Lie superalgebras.} where no external $\you$ is needed.
We leave it as an exercise to the
industrious reader to work out her or his favourite examples.


\begin{thebibliography}{99}
\bibitem{halpkit} M.B. Halpern and E. Kiritsis, Mod. Phys. Lett. A4 (1989)
1373; Erratum {\em ibid.} A4 (1989) 1797.
\bibitem{moretal} A.Y. Morozov, A.M. Perelomov, A.A. Rosly, M.A. Shifman, and
A.V. Turbiner,  Int. Journ. Mod. Phys. A5 (1990) 803.
\bibitem{irrcons} M.B. Halpern, E. Kiritsis, N.A. Obers, M. Porrati, and J.P.
Yamron, Int. Journ. Mod. Phys. A5 (1990) 2275;
M.B. Halpern and N.A. Obers, Int. Journ. Mod. Phys. A6 (1991) 1835,
Nucl. Phys. {\bf B345} (1990) 607;
M.B. Halpern and J.P. Yamron,  Nucl. Phys. {\bf B351} (1991) 333;
A. Giveon, M.B. Halpern, E.B. Kiritsis, and N.A. Obers, Nucl. Phys. {\bf B357}
(1991) 655; \ldots
\bibitem{schratro} S. Schrans and W. Troost, Nucl. Phys. {\bf B345} (1990) 584.
\bibitem{bbss} F.A. Bais, P. Bouwknegt, M. Surridge, and K. Schoutens,  Nucl.
Phys. {\bf B304} (1988) 348; {\bf B304} (1988) 371.
\bibitem{romans} L. Romans, Nucl. Phys. {\bf B352} (1991) 829.
\bibitem{fazam} V.A. Fateev and A.B. Zamolodchikov, Nucl. Phys.
{\bf B280} [FS18] (1987) 644.
\bibitem{zamo} A.B. Zamolodchikov,  Teor. Mat. Fiz. {\bf 65} (1985) 1205.
\bibitem{kris} K. Thielemans, {\sl A} {\em Mathematica$^{TM}$} {\sl Package for
Computing Operator Product Expansions}, Leuven preprint KUL-TF-91/24,
Int. Journ. Mod. Phys. C to be published.
\bibitem{godolwat} P. Goddard, D. Olive, and G. Waterson, Commun. Math. Phys.
{\bf 112} (1987) 591.
\bibitem{liesuper} A. Deckmyn and W. Troost, {\sl Affine Lie superalgebras and
conformal stress tensors}, Leuven preprint KUL-TF-91/16.
\bibitem{morshiftur} A.Y. Morozov, M.A. Shifman, and A.V.
Turbiner, Int. Journ. Mod. Phys. A5 (1990) 2953.
\bibitem{faly} V.A. Fateev and S.L. Lykyanov, Int. Journ. Mod. Phys. A3
(1988) 507.
\bibitem{hullletter} C. Hull, Phys. Lett. {\bf 240B} (1989) 110.
\bibitem{anomgrav} C. Hull, {\sl W-gravity anomalies 1 and 2}, Queen Mary
preprints
QMW/PH/91/2 and QMW/PH/91/3;\\
K. Schoutens, A. Sevrin,
and P. van Nieuwenhuizen, {\sl Quantum $W_3$ gravity in the chiral gauge},
Stony Brook preprint ITP-SB-91-07, Nucl. Phys. {\bf B} to be published.
\bibitem{mieg} J. Thierry-Mieg, Phys. Lett. {\bf 197B} (1987) 368.
\end{thebibliography}
\end{document}